\documentclass{article}%
\usepackage[affil-it]{authblk}
\usepackage{amssymb}
\usepackage{amsmath}
\usepackage{amsfonts}
\usepackage{graphicx}
\usepackage{color}%
\begin{document}

\title{Arithmetic and pseudo-arithmetic billiards}\date{}
\author{P.  Braun}
\affil{Fachbereich Physik, Universit{\"a}t Duisburg-Essen,
45117 Essen, Germany,\\
Institute of Physics, Saint-Petersburg University, 198504  
Saint-Petersburg, Russia}
\maketitle

\begin{abstract}
The arithmetic triangular billiards are classically chaotic but have
Poissonian energy level statistics, in ostensible violation of the BGS
conjecture. We show that the length spectra of their periodic orbits divides
into subspectra differing by the parity of the number of reflections from the
triangle sides; in the quantum treatment that parity defines the reflection
phase of the orbit contribution to the Gutzwiller formula for the energy level
density. We apply these results to all 85 arithmetic triangles and establish
the boundary conditions under which the quantum billiard is \textquotedblleft
genuinely arithmetic\textquotedblright, i. e., has Poissonian level
statistics; otherwise the billiard is "pseudo-arithmetic" and belongs to the
GOE universality class

\end{abstract}

\section{\bigskip Introduction}

Quantum systems whose classical counterpart is chaotic, have level statistics
belonging to one of the Wigner-Dyson universality classes of the random matrix
theory; classically regular systems with more than one degree of freedom have
Poissonian level statistics \cite{Haake10}. This statement, originally the
\textquotedblleft BGS conjecture\textquotedblright\cite{Bohig84,Casat80}, is
now well established on the basis of the semiclassical
approximation \cite{Berry85, Siebe01,Muell05,Muell09} and some essential
properties of the chaotic motion like orbit bunching \cite{Altla09}. The so
called arithmetic systems represent thus a paradox; they have completely
chaotic classical limit but level statistics close to Poissonian
\cite{Auric91,Bolte92,Sarna95,Bogom03,Bogom97,Bolte09}.

Here we concentrate on the triangular arithmetic billiards
\cite{Takeu77,Ninne95,Auric95}. They are formed by geodesics on the Riemann
surface of constant negative curvature and have angles $\pi/l,\pi/m,\pi/n$
where $l,m,n$ are certain integers or infinity; there exist 85 arithmetic
triangles. Consecutively reflecting the triangle in its sides we can tesselate
the complete Riemann surface; the group $T^{\ast}\left(  l,m,n\right)  $
generated by the reflections is the symmetry group of the Hamiltonian of a
particle moving on the tesselated surface. Due to that symmetry, there exists
an infinite amount of the so called Hecke integrals of motion such that the
level repulsion associated with Wigner-Dyson level statistics is absent and
Poissonian statistics arises.

In a more naive approach we can solve the Schr\"{o}dinger equation for the
particle in an isolated triangle imposing Dirichlet or Neumann boundary
conditions on its sides. As demonstrated numerically, only certain
combinations of boundary conditions lead to the spectra with Poissonian level
statistics \cite{Auric95}; we shall then speak of a \textquotedblleft genuinely
arithmetic\textquotedblright, or simply arithmetic,\ quantum billiard as
opposed to the \textquotedblleft pseudo-arithmetic\textquotedblright\ ones
with Wigner-Dyson levels statistics of the GOE class. E. g., if one of the
angles is an odd fraction of $\pi$, the triangle can be genuinely arithmetic
only if the boundary conditions on the adjacent sides of that angle are the
same  \cite{Auric95,Ninne95}. In a problem following from desymmetrization of the solution on the
tesselated Riemann surface, only such boundary conditions occurs.

To completely make peace with BGS, one has to show that the semiclassical
treatment of the arithmetic systems indicates indeed Poissonian statistics. At
a first glance it looks straightforward. A convenient measure of the spectral
statistics is the form factor $K\left(  \tau\right)  $ obtained by the Fourier
transformation of the level-level correlation function \cite{Haake10}. The
diagonal approximation \cite{Berry85} applicable at least at small times,
gives $K\left(  \tau\right)  \approx g\left(  \tau T_{H}\right)  \tau$ where
$T_{H}$ is the Heisenberg time and $g\left(  T\right)  $ is the average action
multiplicity of periodic orbits with period $T$. In a \textquotedblleft
normal\textquotedblright\ system of the GOE universality class $g=2$; on the
other hand, in the arithmetic systems  $g(T)\propto e^{\lambda T/2}/\lambda T$ where $\lambda$ is the Lyapunov constant. Therefore $g\left(  \tau T_{H}\right)  $ grows to
infinity in the limit $\hbar\rightarrow0$ when $T_{H}\rightarrow\infty$ such
that the form factor rises almost vertically from zero at $\tau=0_{+}$ to values comparable with $1$,
as it should be for systems with Poissonian level statistics.

What remained unanswered for some time was why some triangles are genuinely
arithmetic and others only pseudo-arithmetic, since the boundary conditions
don't influence the periodic orbit set. The reason turned out to be the
reflection phase with which the periodic orbit contributes to the Gutzwiller
formula for the level density \cite{Braun10}. Only if the reflection phase is
the same for all orbits comprising every length multiplet, their
contributions to the form factor interfere constructively and the form factor
does behave like $g\left(  \tau T_{H}\right)  \tau$. As shown for the triangle
$\left(  2,3,8\right)  $, the possibility of the constructive interference is
connected with the division of the periodic orbit multiplets into several
classes differing by parities of the number of reflections from the triangle
sides \cite{Ninne95,Braun10}.

In the present paper we extend the investigation to all 85 arithmetic
triangles establishing the orbit length spectra subdivisions and the 
arithmetic boundary conditions. Our reasoning will not be much more
complicated than, say, \textquotedblleft$q_{0}+q_{1}\sqrt{2}$ with rational
$q_{0/1}$ cannot be equal to $\sqrt{3}\left(  q_{0}^{^{\prime}}+q_{1}%
^{^{\prime}}\sqrt{2}\right)  $ with rational $q_{0/1}^{^{\prime}}$ unless all
$q$'s are zero\textquotedblright. We show that three scenarios are realized:

\begin{itemize}
\item The reflection phase of all orbits within a length multiplet is the same
regardless of the boundary conditions, hence the quantum triangle is always
genuinely arithmetic. This is the rarest case found in only four right
triangles $\left(  2,m,n\right)  $ when $m$ and $n$ are both even and not
multiples of each other;

\item The triangle is genuinely arithmetic if certain two sides of the
triangle are both Dirichlet or both Neumann. Observed in 51 triangles;

\item The triangle is genuinely arithmetic if the boundary conditions on all
sides are the same. That scenario is realized in the equilateral triangles and
in triangles whose two angles are odd fractions of $\pi$ or zero. Observed in
30 triangles.
\end{itemize}

\bigskip

\section{Preliminaries}

\subsection{\bigskip Gutzwiller level density, reflection phase, form factor}

The semiclassical theory of the quantum spectral statistics is based on the
Gutzwiller trace formula for the level density \cite{Gutzw90};  for a billiard on pseudosphere it can be written, 
\[
\rho\left(  E\right)  \propto\sum_{\gamma}A_{\gamma}\exp\left(  i\frac
{S_{\gamma}}{\hbar}+i\Phi_{\gamma}\right)
\]
where $\gamma$ are periodic orbits with the action $S_{\gamma}$ and the stability
coefficients $A_{\gamma}$. The phase $\Phi_{\gamma}$ takes into account the
phase gain $\pi$ after every reflection from the sides with the Dirichlet
boundary condition,
\[
\Phi_{\gamma}=\nu_{L}^{\left(  \gamma\right)  }\phi_{L}+\nu_{M}^{\left(
\gamma\right)  }\phi_{M}+\nu_{N}^{\left(  \gamma\right)  }\phi_{N}.
\]
Here $\nu_{L/M/N}^{\left(  \gamma\right)  }$ is the number of visits of the
side $L,M$ or $N$ by the orbit $\gamma$ and $\phi_{L/M/N}$ is $0$ (or $\pi$)
if the boundary condition on the side is Neumann (Dirichlet). Substituting the
Gutzwiller density into the two-point correlation function we obtain, after
the Fourier transform, the form factor as a double sum over orbits with the period
in the interval $t,t+\Delta$ where $\Delta$ is some small time interval,%
\[
K\left(  \tau\right)  \propto\frac{1}{\Delta}\sum_{t<T_{\gamma}<t+\Delta
}A_{\gamma}A_{\gamma^{\prime}}\exp\left[  \frac{i\left(  S_{\gamma}%
-S_{\gamma^{\prime}}\right)  }{\hbar}+i\left(  \Phi_{\gamma}-\Phi
_{\gamma^{\prime}}\right)  \right]  ,\quad t=\tau T_{H}.
\]
Berry's diagonal approximation neglects pairs of orbits with different
actions, or which is the same in a billiard, with different lengths. Denoting
the multiplets of orbits with the same length by $\Lambda$ we can write the
form factor of the diagonal approximation as
\begin{equation}
K_{\mathrm{diag}}\left(  \tau\right)  \propto\sum_{\Lambda}A_{\Lambda}^{2}%
\sum_{\gamma,\gamma^{\prime}\in\Lambda}\left(  \pm1\right)  \label{forfac}%
\end{equation}
where $\left(  \pm1\right)  =\exp\left[  i\left(  \Phi_{\gamma}-\Phi
_{\gamma^{\prime}}\right)  \right]  $. In a \textquotedblleft
normal\textquotedblright\ GOE-system we would have $\gamma^{\prime}=\gamma$ or
$\gamma^{\prime}=\left(  \gamma\right)  _{\mathrm{TR}}$ (\textquotedblleft
TR\textquotedblright=time reversed); both $\gamma$ and $\left(  \gamma\right)
_{\mathrm{TR}}$ have the same number of bumps against any side such that their
reflection phases always cancel. In the arithmetic systems the numbers of
visits $\nu_{L/M/N}^{\left(  \gamma\right)  }$ vary greatly within the length
multiplets and needn't coincide for $\gamma^{\prime},\gamma$ . It is not
obvious why the proclaimed enhancement of the form factor due to the
pathologically high action/length multiplicity is not destroyed by cancelation
of $\left(  \pm1\right)  $ in the inner sum. To put it bluntly, how can the
energy spectrum of an arithmetic triangle be Poissonian unless all its sides
are Neumann?

The question was answered in \cite{Braun10} which used the triangle $\left(
2,3,8\right)  $ as example. It was shown that all orbits belonging to a length
multiplet have the same parity of the number of reflections $\nu_{M}$ from the
side opposite to the angle $\pi/3$ and the same parity of the total number of
reflections $\nu_{L}+\nu_{N}$ from its adjacent sides; the parity of $\nu_{L}$
and $\nu_{N}$ separately is not fixed. Therefore if the boundary condition on
$L$ and $N$ is the same, the reflection phases of $\gamma,\gamma^{\prime}$
belonging to the same length multiplet cancel such that all summands in the
inner sum in (\ref{forfac}) are 1. The form factor then indeed is given by
$g\left(  \tau T_{H}\right)  \tau$ at small $\tau$ , i. e., is
Poissonian-like, and the triangle is genuinely arithmetic. In a
pseudo-arithmetic $\left(  2,3,8\right)  $ the orbit pairs other than the GOE
ones, make mutually cancelling contributions such that the Wigner-Dyson
$K_{\mathrm{diag}}\left(  \tau\right)  \sim2\tau$ is restored.

Below we make a similar purely classical investigation of all 85 arithmetic triangles.

\subsection{Some mathematical reminders}

\begin{itemize}
\item An\emph{\ algebraic number }$\eta$\emph{\ of degree }$n$ is the root of
a polynomial with integer coefficients whose minimal power is $n$; that
polynomial is called the \emph{minimal polynomial }of $\eta$.

\item A \emph{field} $\mathcal{K}$ is a set of objects closed with respect to
addition, subtraction, multiplication and division by a
non-zero. Dropping division we would define a \emph{ring}. Example:
rational numbers form the field $Q$; usual integers form a ring.

\item An \emph{algebraic field} $Q\left(  \eta\right)  $ where $\eta$ is
algebraic, is an extension of $Q$ consisting of the results of the field
operations on the binomials $q_{0}+q_{1}\eta$ with $q_{0},q_{1}$ rational; the
\emph{degree} of $Q\left(  \eta\right)  $ is the power of the minimal
polynomial of $\eta$. Any element of the field $q\left(  \eta\right)  \in
Q\left(  \eta\right)  $ can be uniquely represented by a polynomial,%
\[
q\left(  \eta\right)  =q_{0}+q_{1}\eta+\ldots+q_{n-1}\eta^{n-1}%
\]
where $q_{0},\ldots,q_{n-1}$ are rationals, $n$ is the degree of $\eta$.

\item An \emph{algebraic field }$Q\left(  \eta,\zeta\right)  $%
\emph{\ generated by two algebraic numbers} consists of binomials $q_{0}%
+q_{1}\eta,\quad q_{0}^{\prime}+q_{1}^{\prime}\zeta$ with rational
coefficients, their products and sums of the products. E.g., $Q\left(  \cos
\pi/4,\cos\pi/6\right)  $ consists of the numbers $q\left(  \sqrt{2},\sqrt
{3}\right)  =q_{0}+q_{1}\sqrt{2}+q_{2}\sqrt{3}+q_{3}\sqrt{6}$ and is
equivalent to the field $Q\left(  \cos\pi/12\right)  $.
\end{itemize}

\bigskip It will often be needed to know whether a certain algebraic number
$\zeta$ belongs to the field $Q\left(  \eta\right)  $; in most cases the
answer will be obvious. Note a useful relation concerning the algebraic
numbers $\cos\pi/m$ where $m$ is integer,%
\begin{align}
\cos\frac{\pi}{m}  &  \in Q\left(  \cos\frac{2\pi}{m}\right)  ,\quad m\text{
odd;}\label{cospim}\\
\cos\frac{\pi}{m}  &  \notin Q\left(  \cos\frac{2\pi}{m}\right)  ,\quad
m\text{ even.}\nonumber
\end{align}
The algebraic computer subroutines like $"$\textrm{ToNumberField}$[\zeta
,\eta]"$ of Wolfram Mathematica can be helpful in less transparent situations.

\subsection{M\"{o}bius transformations in Poincar\'e plane}

\bigskip Classical periodic orbits in the arithmetic triangular billiards
consist of pieces of geodesics separated by specular reflections from the
triangle sides. The motion on the Riemann surface is conveniently mapped to
the Poincar\'{e} complex half-plane $\operatorname{Im}z>0$ where the geodesics
are depicted either by circles with the center on the real axis or by straight
lines parallel to the imaginary axis \cite{Bogom97}. After the reflection
$\hat{\Sigma}$ in the triangle side the points $z$ of a geodesic are transformed
into the points $z^{\prime}$ of the mirror-reflected geodesic by the complex
conjugation followed by  M\"{o}bius transformation, $z^{\prime}=\frac
{a\bar{z}+b}{c\bar{z}-a}.$ The matrix of reflection
\[
\Sigma=%
\begin{pmatrix}
a & b\\
c & -a
\end{pmatrix}
\]
is defined up to a sign; it has real elements and $\det\Sigma=-1$. The reflection operation must not be
mixed with its matrix; e. g., $\hat{\Sigma}^{2}$ is identity but $\Sigma^{2}$
can be $+I$ or $-I$. The product of two reflections in the sides adjacent to
the angle $\alpha$ is equivalent to rotation by $2\alpha$ about the crossing
point. Transformation of the geodesics after such a rotation is  M\"{o}bius
transformation whose matrix is a product of the reflection matrices; its
determinant is 1 and its trace coincides with $2\cos\alpha$, up to a sign,

Any periodic orbit in a triangle can be encoded by the list of its $\nu$
consecutive reflections; the trace of the product $W$ of the respective
reflection matrices (\textquotedblleft the orbit trace\textquotedblright\ for
short) is connected with the orbit length $l$ by,%
\begin{align*}
2\cosh\frac{l}{2} &  =\left\vert \operatorname*{Tr}W\right\vert ,\quad
\nu\text{ even;}\\
2\sinh\frac{l}{2} &  =\left\vert \operatorname*{Tr}W\right\vert ,\quad
\nu\text{ odd.}%
\end{align*}
Many different orbits can have the same trace which leads to the length
multiplet formation. On the other hand, two orbits, $\gamma_{1}$ with an even
number of reflections, and $\gamma_{2}$ with an odd one, can have equal length
only if%
\begin{equation}
\left(  \operatorname*{Tr}W_{\gamma_{1}}\right)  ^{2}-\left(
\operatorname*{Tr}W_{\gamma_{2}}\right)  ^{2}=4.\label{nudegener}%
\end{equation}
This is an additional constraint which either excludes formation of the length
multiplets with mixed parity of $\nu$ or reduces their number to a proportion
exponentially small in the limit of large orbit lengths. Complete
investigation of the exceptional situations when (\ref{nudegener}) is
fulfilled, can be found for the triangle $\left(  2,3,8\right)  $ in the
archived version of \cite{Braun10}. Since our interest lies in the quantum
level statistics which is not influenced by the exceptional orbits, we do not
extend this investigation to other arithmetic triangles. Numerical simulations
confirm that the multiplets with mixed parity of $\nu$ are indeed either
extremely rare or absent in all arithmetic triangles. An immediate physical
consequence is that all arithmetic triangles with the Dirichlet boundary
conditions on every side, are genuinely arithmetic.

\subsection{Approximate length degeneracy of orbits with even and odd number
of reflections}

\begin{figure}[tbh]
\includegraphics[width=0.5\textwidth]{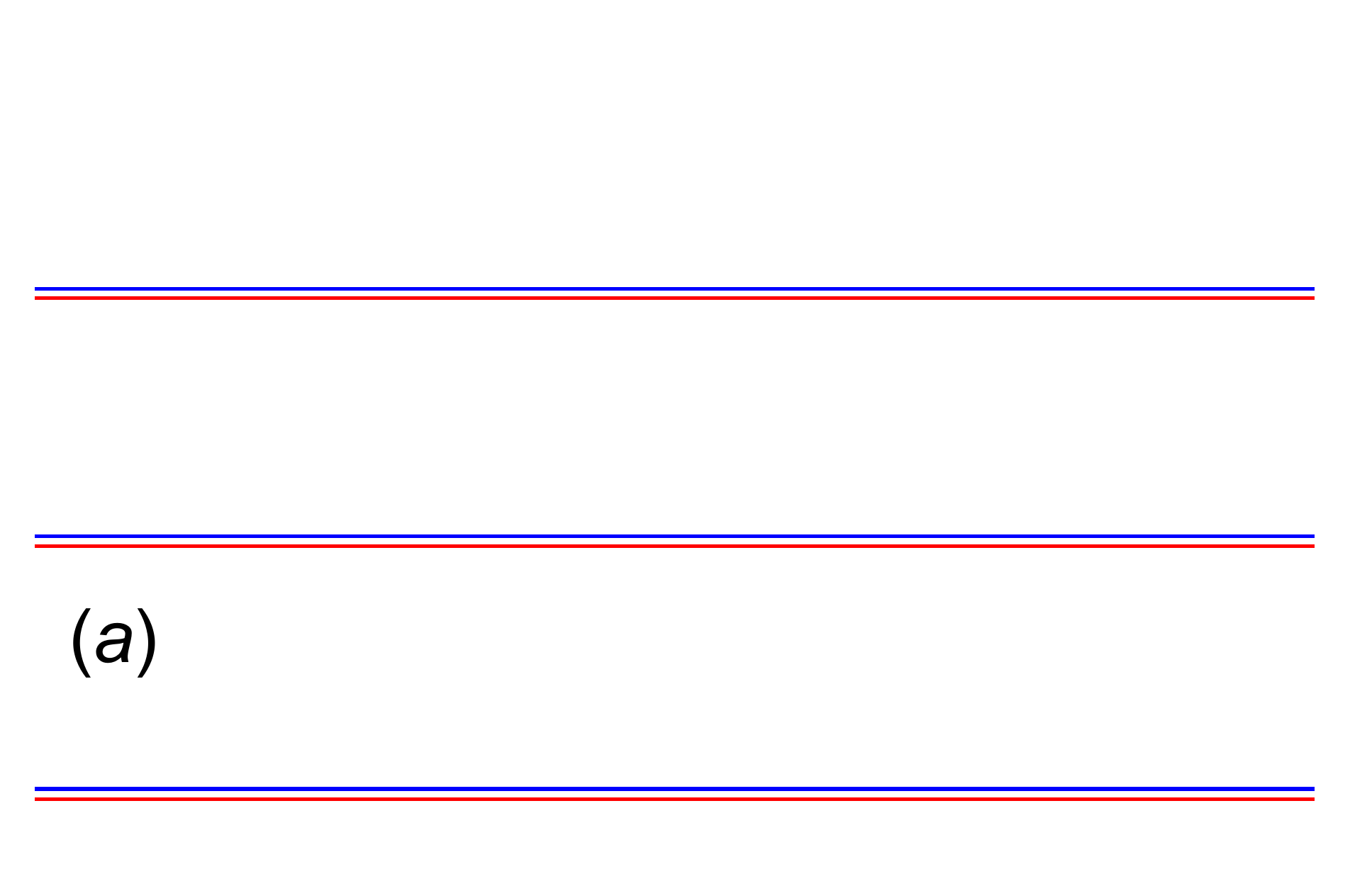} \hfill
\includegraphics[width=0.5\textwidth]{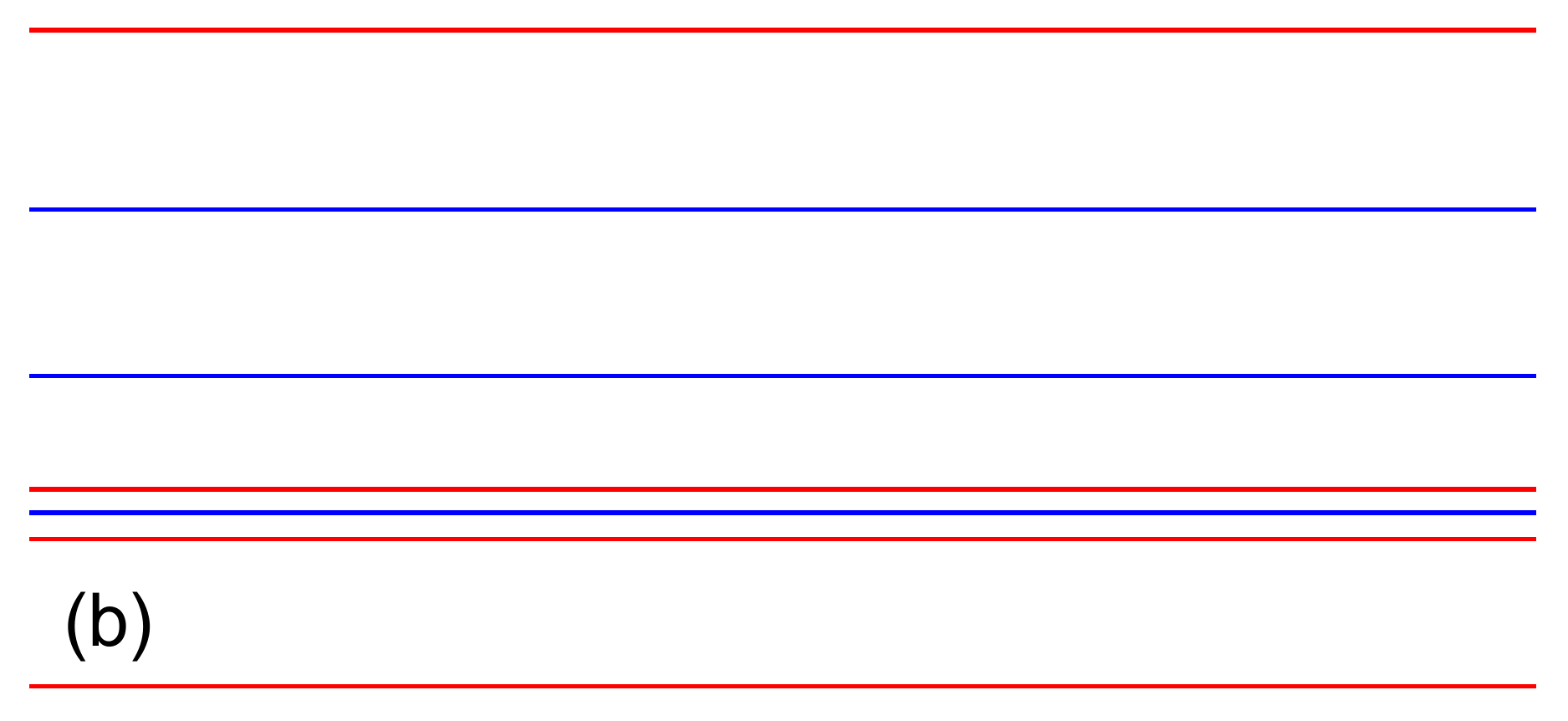} \newline%
{
\parbox[b]{0.3\textwidth}{
} \hfill
\parbox[b]{0.5\textwidth}{
}
\caption{\label{togeta} Stretch of PO  length spectra of a) Artin's billiard $(2,3,\infty)$  and b) the billiard $(2,3,8)$.  Lengths of orbits with even and odd number of reflections are shown by blue and red lines respectively }
}
\end{figure}

Unlike exact coincidence, systematic approximate equality of lengths
$l_{e},l_{o}$ of the orbits with even (e) and odd (o) number of reflections $\nu$ does occur in some
billiards. Namely, it happens when the corresponding traces coincide exactly,
$\operatorname*{Tr}W_{e}=\operatorname*{Tr}W_{o}$; the necessary condition of
the $eo-$trace degeneracy is given below in the end of Section \ref{odd_no_ref}. The orbit
lengths exist then in doublets with the spacing $\Delta l_{oe}=l_{o}%
-l_{e}\approx4e^{-l}$ which is to be compared with the much greater spacing between
the length of orbits in the absence of the trace degeneracy, $\Delta
l_{av}=\mathrm{const\times}e^{-l/2}$. The well-known example is Artin's
billiard $\left(  2,3,\infty\right)  $; a short stretch of its length spectrum
around $l=8$ is shown in Fig.\ref{togeta}a. For comparison, we show a similar
plot of the lengths of orbits of the billiard $\left(  2,3,8\right)  $, see
Fig.\ref{togeta}b, in which the $eo-$trace degeneracy is forbidden by the
arithmetic considerations. 

The importance of the approximate $eo$ length degeneracy becomes obvious when
we consider the off-diagonal contributions $\propto\cos\frac{\Delta S}{\hbar}$
of the $eo-$doublets  to the form factor $K\left(  \tau\right)  $.It is easy to show that in the semiclassical regime 
the phase difference $\Delta S/\hbar\ $ is much larger than 1 when the time is smaller than the 
Ehrenfest time $\tau_{E}=T_{E}/T_{H}$.  For   $\tau>\tau_{E}$ the phase difference exponentially fast tends to zero  such that the $eo$-splitting can be neglected; the diagonal approximation must therefore to be reformulated.
In the crossover region around $\tau_{E}$ the behavior of  $K(\tau)$ is expected to be complicated; cf. the exact $K\left(  \tau\right) $ for  Artin's billiard obtained in \cite{Bogom96}.

 The \textquotedblleft normal\textquotedblright%
\ off-diagonal contributions connected with the off-diagonal orbit pairs with
non-coinciding $\operatorname*{Tr}W$ become significant at  times $\tau
\gtrsim$ $2\tau_{E}$; systematic contributions of that type are expected to
arise from the pairs of orbit-partners consisting of approximately the same
pieces traversed in a different order and, perhaps, with different sense\cite{Altla09}.  

\section{Method}

\subsection{Reflection matrices; original and mirror triangles}

All but 5 arithmetic triangles are right or can be reduced to the right ones
with non-equal legs by desymmetrization. Therefore we start with the triangles
$\left(  2,m,n\right)  ,\quad m\ne n, $ such that $L$ is the hypotenuse of the
triangle and $N,M$ are its two legs. The vertices formed by the crossing of
$LM,\,MN,\,LN$ will be denoted $O,Q$ and $P$; the respective angles will be
$\alpha=\frac{\pi}{n},\,\pi/2$ and $\beta=\frac{\pi}{m}$. It will be
convenient to direct the side $M$ along the imaginary axis and choose
$z_{Q}=i$; then the two other vertices will be
\begin{align*}
z_{O}  &  =i\frac{\cos\beta+\rho}{\sin\alpha},\quad z_{P}=\frac{\rho
+i\sin\beta}{\cos\alpha},\\
\rho &  \equiv\sqrt{\cos^{2}\beta+\cos^{2}\alpha-1}.
\end{align*}
\begin{figure}[tbh]
\includegraphics[width=0.29\textwidth]{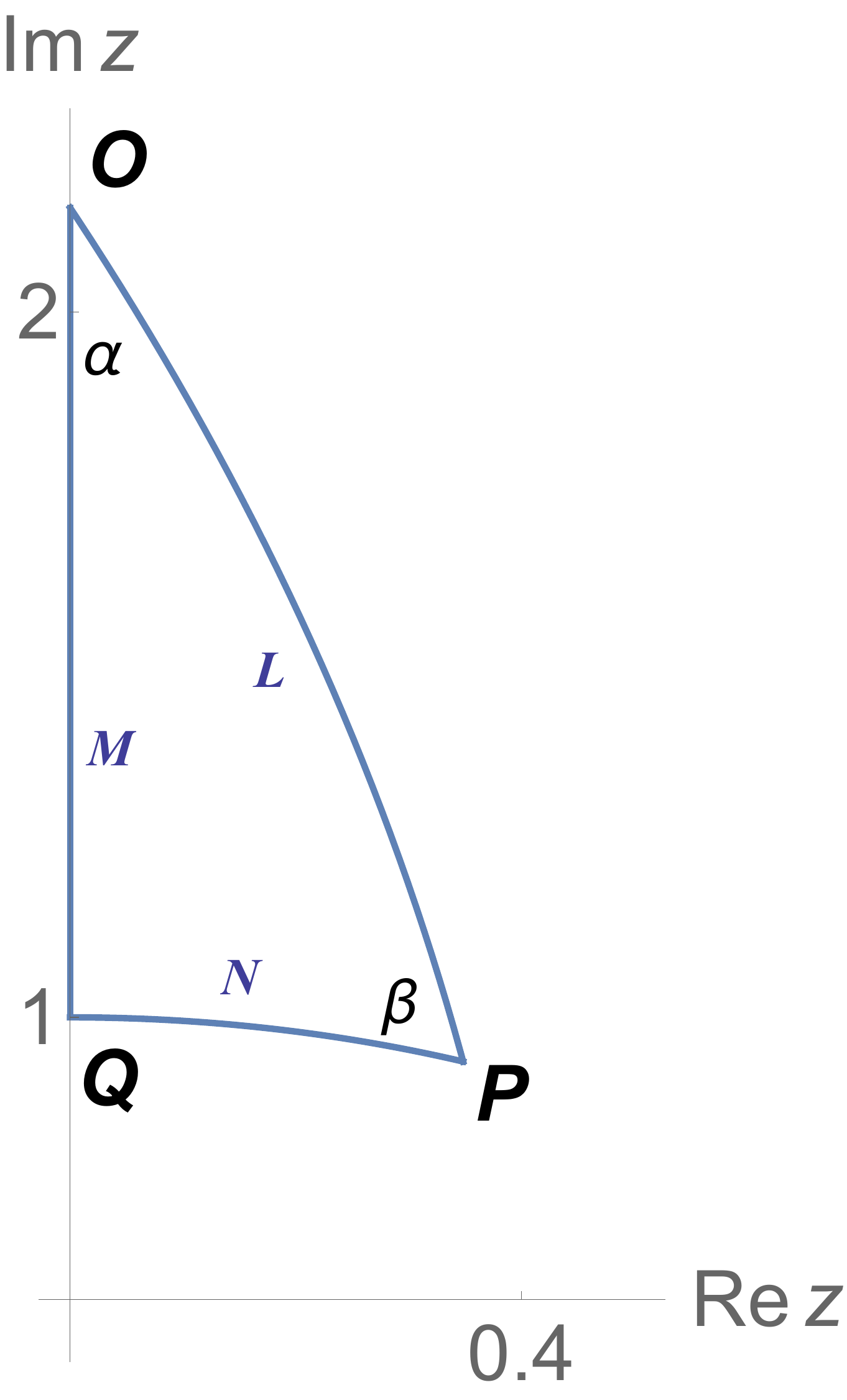} \hfill
\includegraphics[width=0.4\textwidth]{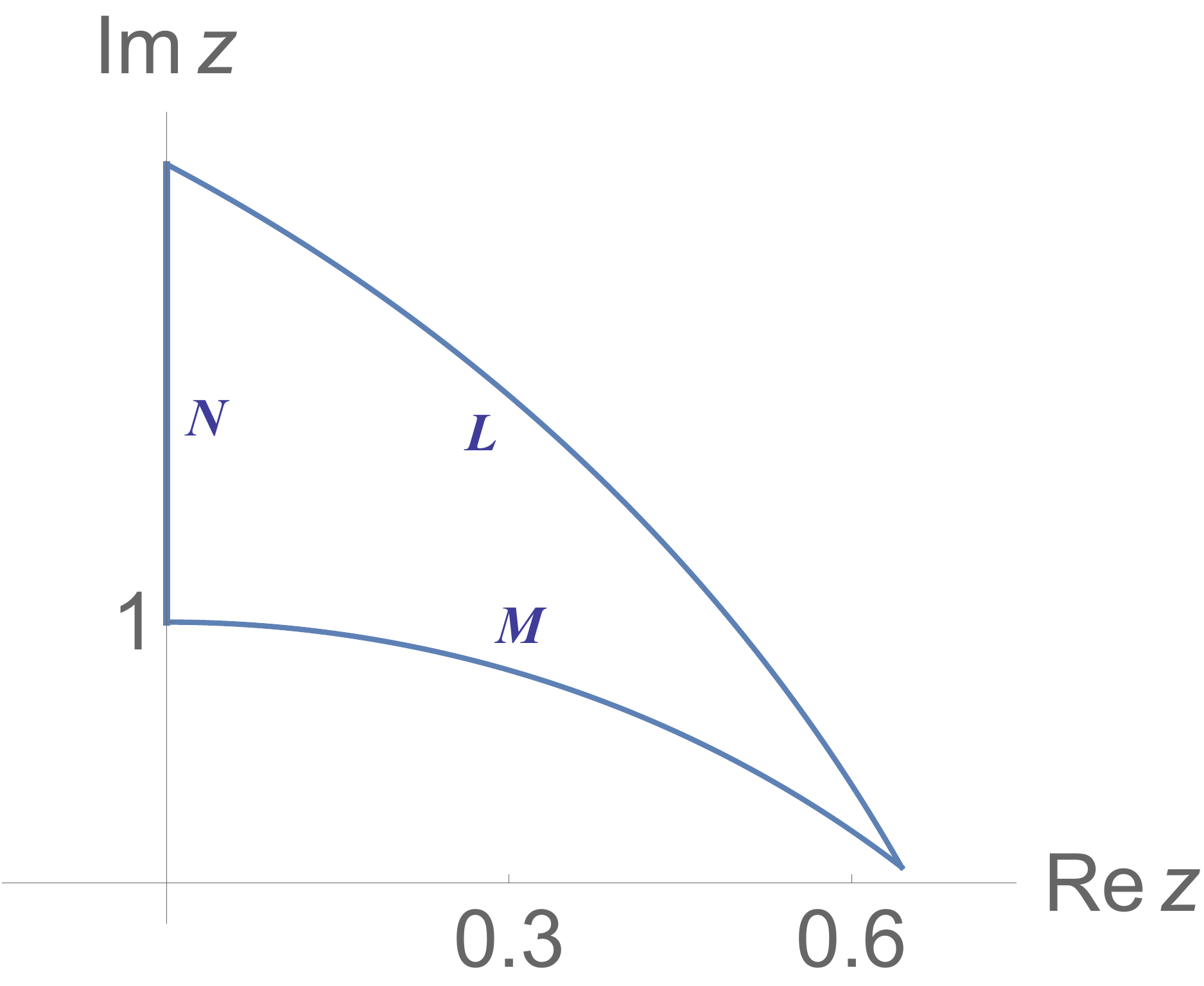} \newline%
\parbox[b]{0.3\textwidth}{\caption{Right triangle in Poincar\'{e} half-plane}
\label{Fig1}} \hfill
\parbox[b]{0.5\textwidth}{\caption{Same triangle reflected in the bisector of the right angle}\label{Fig2}}
\end{figure}
The M\"{o}bius matrices of reflections in the sides $M,N,L$ are
\begin{equation}
\Sigma_{M}=%
\begin{pmatrix}
1 & 0\\
0 & -1
\end{pmatrix}
,\quad\Sigma_{N}=%
\begin{pmatrix}
0 & 1\\
1 & 0
\end{pmatrix}
,\quad\Sigma_{L}\left(  \alpha,\beta\right)  =%
\begin{pmatrix}
-\cos\alpha & \cos\beta+\rho\\
\cos\beta-\rho & \cos\alpha
\end{pmatrix}
.
\end{equation}
Their products describe rotations by $\pi,2\beta$ and $2\alpha$ about $Q,P$
and $O$,%
\begin{align}
R_{Q}  &  =\Sigma_{M}\Sigma_{N}=%
\begin{pmatrix}
0 & 1\\
-1 & 0
\end{pmatrix}
,\label{Elemrot}\\
R_{P}\left(  \alpha,\beta\right)   &  =\Sigma_{N}\Sigma_{L}=%
\begin{pmatrix}
\cos\beta-\rho & \cos\alpha\\
-\cos\alpha & \cos\beta+\rho
\end{pmatrix}
\,,\nonumber\\
R_{O}\left(  \alpha,\beta\right)   &  =\Sigma_{L}\Sigma_{M}=%
\begin{pmatrix}
-\cos\alpha & -\cos\beta-\rho\\
\cos\beta-\rho & -\cos\alpha
\end{pmatrix}
.\nonumber
\end{align}

The code $\mathbb{R}$ of the orbit with an even number of reflections can be
written as a sequence of the elementary rotations. The trace of the corresponding orbit
matrix  $R$ belongs to a ring with integer coefficients generated by $2\cos
\pi/l,\,\,2\cos\pi/m,\,\,2\cos\pi/n,$ \cite{Takeu77}; in fact in all cases one of the
generators can be dropped because it is either integer (case $l=2,3,\infty)$
or coincides with one of the other two generators (symmetric triangles). For
our purposes a weaker statement is sufficient that the traces are elements of
the field%
\begin{equation}
\operatorname*{Tr}R\mathbb{\in}Q\left(  \cos\pi/m,\cos\pi/n\right)  \equiv
K_{T}.\label{FieldKt}%
\end{equation}
We shall show that in the majority of the arithmetic triangles this field can
be divided into four or two sets which have no common elements except zero;
membership in a particular set is defined by parity of the number of
elementary rotations composing $\mathbb{R}$.

Simultaneously with $\left(  2,m,n\right)  $ we consider its mirror twin
$\left(  2,n,m\right)  $ directingt its leg $N$ opposite to $\pi/n$ along
the imaginary axis in the Poincar\'{e} half-plane (Fig.~2). The corresponding
periodic orbits in the two triangles are built of the same sequences of
reflections $\hat{\Sigma}_{L},\hat{\Sigma}_{M},\hat{\Sigma}_{N}$ in the sides
opposite to the angles $\pi/l,\pi/m,\pi/n$; the related M\"{o}bius
transformations will be given by the matrices $\Sigma_{L}\left(  \beta
,\alpha\right)  ,\,\Sigma_{N},\,\Sigma_{M}$ introduced above. Therefore that
\textquotedblleft mirror approach\textquotedblright\ amounts to the
substitutions $\alpha\leftrightarrows\beta,M\leftrightarrows N$ in the matrix products.

\subsection{Even number of reflections: alternatives for the matrix traces}

The matrix of an orbit with even number of reflections is a rotation
matrix and, in the  representation we have chosen, can be one of the two types. The first
one is the  Fuchsian matrix \cite{Takeu77,Bogom97,Bolte09} whose general
form is,%
\begin{equation}
R^{\left(  I\right)  }=%
\begin{pmatrix}
x_{0}+x_{1}\sqrt{a} & x_{2}\sqrt{b}+x_{3}\sqrt{ab}\\
x_{2}\sqrt{b}-x_{3}\sqrt{ab} & x_{0}-x_{1}\sqrt{a}%
\end{pmatrix}
\label{matr gamma}%
\end{equation}
where $x_{0},\ldots,x_{3},a>0,b>0$ are real numbers belonging to some
algebraic field $K$; the determinant of the matrix is $1$. The Fuchsian
matrices form a group: the product of two matrices $R_{1}^{\left(  I\right)
}R_{2}^{\left(  I\right)  }$ with the same $K,a,b$ but different $x$ is also
Fuchsian with the same $K,a,b$.
 Our elementary rotation matrix $R_{P}$ is
Fuchsian with%
\begin{align}
K &  =Q\left(  \cos^{2}\alpha,\cos\beta\right)  =Q\left(  \cos2\alpha
,\cos\beta\right)  ;\label{matr y}\\
a &  =\cos^{2}\beta+\cos^{2}\alpha-1=\rho^{2},\quad b=\cos^{2}\alpha/\rho
^{2}\nonumber
\end{align}
and $x_{0}=\cos\beta,\quad x_{1}=-1,\quad x_{2}=0,\quad x_{3}=1$; consequently
any power of $R_{P}$ is also of the type $R^{\left(  I\right)  }$ with $K,a,b$
given in (\ref{matr gamma}). The field $K$ is either a subfield of the field $K_T$ defined in (\ref{FieldKt}) or coincides with it.

If we multiply  $R^{\left(  I\right)  }$ by the elementary rotation
matrix $R_{Q}$ we get a different-looking creature,
\begin{equation}
R^{\left(  II\right)  }=R^{\left(  I\right)  }R_{Q}=%
\begin{pmatrix}
-x_{2}\sqrt{b}-x_{3}\sqrt{ab} & x_{0}+x_{1}\sqrt{a}\\
-x_{0}+x_{1}\sqrt{a} & x_{2}\sqrt{b}-x_{3}\sqrt{ab}%
\end{pmatrix}
.\label{defk}%
\end{equation}
The left multiplication of $R^{\left(  II\right)  }$ also produces a matrix of the
type $R^{\left(  II\right)  }$ with $x_{1}\rightarrow-x_{1},x_{2}%
\rightarrow-x_{2}$. Multiplication of $R^{\left(  II\right)  }$ by $R_{Q}$
produces $R^{\left(  I\right)  }$; in other words $R_{Q}$ toggles the matrix
type between $I$ and $II.$   The  multiplication table for the types can be symbolically written,%
\begin{equation}
\label{type_multi}
I\times I=I,\quad II\times II=I,\quad II\times I=II,\quad I\times II=II.
\end{equation}

The elementary rotation $R_{O}$ is of the type $II$ with $x_{0}=-\cos
\beta,x_{1}=-1,x_{2}=0,x_{3}=-1$ such that multiplication of the  orbit matrix by $R_{O}$, same as
by $R_{Q}$,  toggles the matrix type between $I$ and $II$. Consequently, if
the code consists of an even total number of the elementary rotations
$R_{O}=\Sigma_{M}\Sigma_{L}$ and $R_{Q}=\Sigma_{M}\Sigma_{N}$ coinciding with
the number of reflections $\nu_{M}$ from the side $M$, its matrix will belong
to the type $R^{\left(  I\right)  }$, otherwise it is $R^{\left(  II\right)
}$. Since the total number of reflections $\nu$ is assumed even we can just as
well say that the type is $R^{\left(  I\right)  }$ if $\nu_{N}+\nu_{L}=\nu
-\nu_{M}$ is even, and $R^{\left(  II\right)  }$ if it is odd.

A matrix $R^{\left(  II\right)  }$ could be $R^{\left(  I\right)  }$ in
disguise differing only by a similarity transformation. However we can compare
the corresponding traces which are basis independent,%
\[
\operatorname{Tr}R^{\left(  I\right)  }=2x_{0},\quad\operatorname{Tr}%
R^{\left(  II\right)  }=-2x_{3}\sqrt{ab}.
\]
Whereas $\operatorname{Tr}R^{\left(  I\right)  }$ belongs to the field
$K=Q\left(  \cos2\alpha,\cos\beta\right)  $, the trace $\operatorname{Tr}%
R^{\left(  II\right)  }$ does so if and only if $\sqrt{ab}=\cos\alpha$ is
contained in $K$; that field coincides then with the field $K_{T}$. If it doesn't, $K$ is a subfield of $K_T$ and the ``coset''\footnote{The field contains zero and is therefore not a group with respect to multiplication, hence``coset'' in quotation  marks} $\cos\alpha\,\,K$  has no non-zero common elements with
$K$. The traces of $R^{\left(  I\right)  }$ and $R^{\left(
II\right)  }$ cannot then be equal, consequently the corresponding periodic orbits cannot
have equal lengths.

We can repeat these arguments for the mirror-reflected triangle and get an
alernative division of the rotation matrices into two types. The first one
is given by the Fuchsian matrices $R^{\left(  I^{\prime}\right)  }$  with%
\begin{align*}
K^{\prime} &  =Q\left(  \cos\alpha,\cos2\beta\right)  ,\\
a^{\prime} &  =\cos^{2}\beta-\sin^{2}\alpha,\quad\sqrt{a^{\prime}b^{\prime}%
}=\cos\beta;
\end{align*}
the elementary rotation matrix $R_{O}=\Sigma_{M}\Sigma_{L}$ and all its powers
are of the type $R^{\left(  I^{\prime}\right)  }$. Multiplication by $R_{Q}$ and $R_{P}$
changes the matrix type to $II^{\prime}$ analogous to (\ref{defk}) but with $K,a,b$
replaced by $K^{\prime},a^{\prime},b^{\prime}$; the multiplication table for the primed types is similar to (\ref{type_multi}).
Consequently the rotation matrix is of the type $I^{\prime}$ if its code
contains even total number of the elementary rotations $R_{Q}=\Sigma
_{M}\Sigma_{N}$ and $R_{P}=\Sigma_{N}\Sigma_{L}$ coinciding with the number of
reflections $\nu_{N}$ from the side $N$; otherwise its type is $II^{\prime}$.
The corresponding traces look like%
\[
\operatorname{Tr}R^{\left(  I^{\prime}\right)  }=2x_{0}^{\prime}%
,\quad\operatorname{Tr}R^{\left(  II^{\prime}\right)  }=-2x_{3}^{\prime}%
\sqrt{a^{\prime}b^{\prime}}%
\]
where $x^{\prime},a^{\prime},b^{\prime}$ belong to $K^{\prime}$ as well as
$\operatorname{Tr}R^{\left(  I^{\prime}\right)  }$.  The trace
$\operatorname{Tr}R^{\left(  II^{\prime}\right)  }$ will belongs to
$K^{\prime}$ if and only if $\cos\beta\in K^{\prime}$; if it doesn't the
orbits with $\nu_{N}$ even and odd cannot have coinciding lengths.  

Combining the two approaches we get the following alternatives: The fields
$Q\left(  \cos2\alpha,\cos\beta\right)  $ and $Q\left(  \cos\alpha,\cos
2\beta\right)  $ can coincide (e.g., if $\alpha=\beta$) or not; $\cos\alpha$
(resp. $\cos\beta$) can belong to $Q\left(  \cos2\alpha,\cos\beta\right)  $
(resp. $Q\left(  \cos\alpha,\cos2\beta\right)  $) or not. That gives at most
four algebraic types of traces and correspondingly four types of the length
multiplets of periodic orbits with even number of reflections.

\subsection{\bigskip\ Orbits with odd number of reflections}
\label{odd_no_ref}

In the first approach we write the code with an odd number of symbols as
$\hat{\Sigma}_{M}\mathbb{\hat{R}}$ where $\mathbb{\hat{R}}$ is a product of
rotations, and the corresponding M\"{o}bius matrix as $W=\Sigma_{M}R$; note
that $\operatorname*{Tr}W=R_{11}-R_{22}$. The rotational body $R$ contains
$\nu_{M}-1$ reflections in $M$ and looks like (\ref{matr gamma}) or
(\ref{matr y}) depending on parity of $\nu_{M}-1$,
\begin{align*}
\operatorname*{Tr}W &  =2x_{1}\sqrt{a}\in\rho Q\left(  \cos2\alpha,\cos
\beta\right)  ,\quad\nu_{M}-1\quad\text{even,}\\
\operatorname*{Tr}W &  =-2x_{2}\sqrt{b}\in\rho\cos\alpha\,Q\left(  \cos
2\alpha,\cos\beta\right)  ,\quad\nu_{M}-1\quad\text{odd;}%
\end{align*}
we remind that $\rho=\sqrt{\cos^{2}\beta+\cos^{2}\alpha-1}$. In the mirror
approach we write the code as $\hat{W}=\hat{\Sigma}_{N}\mathbb{\hat{R}}$ and
the corresponding matrix as $W=\Sigma_{M}^{\prime}R\,$where $R^{\prime}\,$is
obtained by the substitutions $\hat{\Sigma}_{L}\rightarrow\Sigma_{L}\left(
\beta,\alpha\right)  ,\quad\hat{\Sigma}_{M/N}\rightarrow\Sigma_{N/M}$; the
orbit trace is again the difference of the diagonal elements of $\mathbb{R}%
^{\prime}\,$. Considering that $\rho\left(  \alpha,\beta\right)  =\rho\left(
\beta,\alpha\right)  $ we get
\begin{align*}
\operatorname*{Tr}W &  \in\rho Q\left(  \cos\alpha,\cos2\beta\right)
,\quad\nu_{N}-1\quad\text{even;}\\
\operatorname*{Tr}W &  \in\rho\cos\beta\,Q\left(  \cos\alpha,\cos
2\beta\right)  ,\quad\nu_{N}-1\quad\text{odd.}%
\end{align*}
Therefore the arithmetic types of the orbit traces  are obtained from those of its
rotational part by multiplication by $\rho$; 
 again we can have at most four additional types of the
length multiplets; they all belong to the set $\rho\times  Q_T$. It follows that the eo-trace degeneracy and the Artin-like doublet structure in the PO length spectrum can exist only if  $\rho$  belongs to $Q_T$.  
E. g., for Artin's billiard $(2,3,\infty)$ the field $Q_T$ coincides with the field of rational numbers Q which contains $\rho=1/2$. On the other hand, for the billiard $(2,3,8)$ we have 
\begin{eqnarray}
Q_T=Q\left(\cos \frac \pi 8\right) =Q\left(\sqrt{2+\sqrt 2}\right),\nonumber\\
\rho=\sqrt{\cos^2 \frac \pi 8 + \cos^2 \frac \pi 3 -1}=\frac 1 2 \sqrt{\sqrt{2}-1}\notin Q_T \nonumber
\end{eqnarray}
Correspondingly the doublet structure is present in the first example and  absent in the second one.

\section{Classification of arithmetic triangles}

Next we study the arithmetic triangles grouping them according to the three
scenarios mentioned; the number of types refers to orbits with fixed parity of
$\nu$.

\subsection{Group I: Four trace types; triangles $(2,n,m)$ with $m,n$ even and
$n$ non-divisible by $m$ [4 systems]}

The group consists of the right triangles $\left(  2,4,6\right)  ,\left(
2,4,10\right)  ,\left(  2,4,18\right)  ,\left(  2,6,8\right)  $. Table I
contains the fields encountered in the two approaches for these
systems.\begin{table}[ptb]%
\begin{tabular}
[c]{|l|l|l|l|l|}\hline
$\alpha$ & $\beta$ & $K$ & $K^{\prime}$ & $\rho$\\\hline
$\frac{\pi}{6}$ & $\frac{\pi}{4}$ & $Q\left(  \sqrt{2}\right)  $ & $Q\left(
\sqrt{3}\right)  $ & $\frac{1}{2}$\\\hline
$\frac{\pi}{10}$ & $\frac{\pi}{4}$ & $Q\left(  \sqrt{5},\sqrt{2}\right)  $ &
$Q\left(  \cos\frac{\pi}{10}\right)  $ & $\sqrt{\frac{\sqrt{5}+1}{8}}$\\\hline
$\frac{\pi}{18}$ & $\frac{\pi}{4}$ & $Q\left(  \cos\frac{\pi}{9}\right)
+\sqrt{2}Q\left(  \cos\frac{\pi}{9}\right)  $ & $Q\left(  \cos\frac{\pi}%
{18}\right)  $ & $\sqrt{\frac{1}{2}\cos\frac{\pi}{9}}$\\\hline
$\frac{\pi}{8}$ & $\frac{\pi}{6}$ & $Q\left(  \sqrt{2},\sqrt{3}\right)  $ &
$Q\left(  \sqrt{2+\sqrt{2}}\right)  $ & $\frac{\sqrt{\sqrt{2}+1}}{2}$\\\hline
\end{tabular}
\parbox[t]{0.87\textwidth}
{\caption{Algebraic fields associated with triangles of Group I}}
\end{table}
In all cases the fields $K,K^{\prime}$ do not coincide; $\cos\alpha$ does not
belong to $K$ and $\cos\beta$ does not belong to $K^{\prime}$; that gives the
division of the field $K_{T}$ into four non-overlapping subsets. Consequently
the length multiplets can be of four types depending on parity of $\nu_{M}$
and $\nu_{N}$; parity of $\nu_{L}$ is fixed by the condition that the overall
number of reflections $\nu$ is even or odd. Within a degenerate length
multiplet parity of all three numbers $\nu_{L},\nu_{M},\nu_{N}$ is fixed,
hence all orbits of the multiplet acquire the same reflection phase modulo
$2\pi$. The quantum energy spectrum is thus always be Poissonian and the
triangle is genuinely arithmetic regardless of the boundary conditions.

Let us look in more detail, e. g., at the triangle $\left(  2,4,10\right)  $
starting with $\nu$ even; denote $\kappa=2\cos\pi/10=\sqrt{\frac{5+\sqrt{5}%
}{2}}.$ If $\nu_{M}$ is even the orbit traces will have the structure
$q\left(  \sqrt{5,}\sqrt{2}\right)  =$ $q_{0}+q_{1}\sqrt{2}+\sqrt{5}\left(
q_{2}+q_{3}\sqrt{2}\right)  $ where $q_{0\ldots3}$ are rationals; with
$\nu_{M}$ odd they are $\kappa\,q\left(  \sqrt{5,}\sqrt{2}\right)  $. In the
mirror approach the traces have structure $q\left(
\kappa\right)  =q_{0}^{^{\prime}}+q_{1}^{^{\prime}}\kappa+\sqrt{5}\left(
q_{2}^{^{\prime}}+q_{3}^{^{\prime}}\kappa\right)  $ if $\nu_{N}$ is even and
$\sqrt{2}q\left(  \kappa\right)  $ if $\nu_{N}$ is odd. Now, when both
$\nu_{M}$ and $\nu_{N}$ are even (and consequently $\nu_{L}$ is also even) the
trace is simultaneously $q\left(  \sqrt{5,}\sqrt{2}\right)  $ and $q\left(
\kappa\right)  $, i.e., must be $q\left(  \sqrt{5}\right)  =q_{0}+q_{1}%
\sqrt{5}$. The traces structure for other combinations of parities of
$\nu_{M/N}$ for $\nu$ even is given in Table 2 (e=even, o=odd); these results
can be seen as a refinement of the general formula (\ref{FieldKt}). For the
orbits with $\nu$ odd the additional common factor $\rho=\sqrt{\frac{\sqrt
{5}+1}{8}}$ appears; remembering that the rotational body of the code contains
$\nu_{M}-1$ reflections (resp. $\nu_{N}-1$ reflections in the mirror approach)
we get an analogous Table 3. It can be checked that $\rho$ does not belong either 
to $Q$ or to $Q'$, hence there is no Artin's-like approximate $eo$ orbit length degeneracy.

\begin{table}[tbh]%
\begin{tabular}
[c]{|l|l|l|l|}\hline
$\nu_{M}$ & $\nu_{N}$ & $\nu_{L}$ & Trace\\\hline
e & e & e & $q\left(  \sqrt{5}\right)  $\\\hline
e & o & o & $q\left(  \sqrt{5}\right)  \sqrt{2}$\\\hline
o & e & o & $q\left(  \sqrt{5}\right)  \kappa$\\\hline
o & o & e & $q\left(  \sqrt{5}\right)  \kappa\sqrt{2}$\\\hline
\end{tabular}
\hfill%
\begin{tabular}
[c]{|l|l|l|l|}\hline
$\nu_{M}$ & $\nu_{N}$ & $\nu_{L}$ & Trace\\\hline
o & o & o & $q\left(  \sqrt{5}\right)  \rho$\\\hline
o & e & e & $q\left(  \sqrt{5}\right)  \rho\sqrt{2}$\\\hline
e & o & e & $q\left(  \sqrt{5}\right)  \rho\kappa$\\\hline
e & e & o & $q\left(  \sqrt{5}\right)  \rho\kappa\sqrt{2}$\\\hline
\end{tabular}
\parbox[t]{0.47\textwidth}{\caption{Triangle $(2,4,10)$, even $\nu$}
\label{1stprim_a}} \hfill\parbox[t]{0.47\textwidth}{\caption{Triangle $(2,4,10)$, odd $\nu$}\label{1stprim_b}}\end{table}

.

\subsection{Group II: Two algebraic types [51 systems]}

This is the largest and fairly heterogeneous group.

\subsubsection{ Non-symmetric right triangles $\left(  2,m,n\right)  $ with
$m$ odd or infinite, $n$ even}

\label{II_1} These are the triangles $\left(  2,3,n\right)  $,
$n=8,10,12,14,16,18,24,30;$ $\left(  2,5,n\right)  $, $n=4,6,8,10,20,30;$
$\left(  2,7,4\right)  ,\left(  2,7,14\right)  ,\left(  2,9,18\right)
,\left(  2,15,30\right)  ,$ $\left(  2,\infty,4\right)  ,\left(
2,\infty,6\right)  $.

These triangles have properties similar to the previously investigated
$(2,3,8)$. For all of them $\cos\alpha\notin Q\left(  \cos2\alpha,\cos
\beta\right)  $ but $\cos\beta\in Q\left(  \cos\alpha,\cos2\beta\right)  $.
One can easily check it with the help of (\ref{cospim}) considering that
$\alpha=\pi/n$ with $n$ even and $\beta=\pi/m$ with $m$ odd. It follows that
the trace field $K_{T}$ is divided into the subfield $K=Q\left(  \cos
2\alpha,\cos\beta\right)  $ and its \textquotedblleft coset\textquotedblright%
\ $\cos\alpha\,K$; the mirror approach doesn't produce an alternative division of $K_T$. The
traces of orbits with an even total number of reflections~$\nu$ belong thus
either to $K$ or to $\cos\alpha\,K$ when $\nu_{M}$ is even and odd,
respectively. Therefore the length multiplets cannot contain orbits with
different parity of $\nu_{M}$ and $\nu_{L}+\nu_{N}$; individual parities of
$\nu_{L},\nu_{N}$ are not fixed. The reflection phases of orbits within the
length multiplets will be equal only if the boundary conditions on the
hypotenuse $L$ and the leg $N$ adjacent to the angle $\beta$ coincide; only
then the quantum level statistics will be Poissonian and the triangle
genuinely arithmetic.

For the orbits with odd $\nu$ the traces belong to $\rho\cos\alpha Q\left(
\cos2\alpha,\cos\beta\right)  $ or $\rho Q\left(  \cos2\alpha,\cos
\beta\right)  ,$ when $\nu_{M}$ is even or odd, respectively; conclusions on
parity of the number of reflections  of the orbits in the length multiplets and the quantum
level statistics remain the same.

\begin{figure}[tbh]
\includegraphics[width=0.5\textwidth]{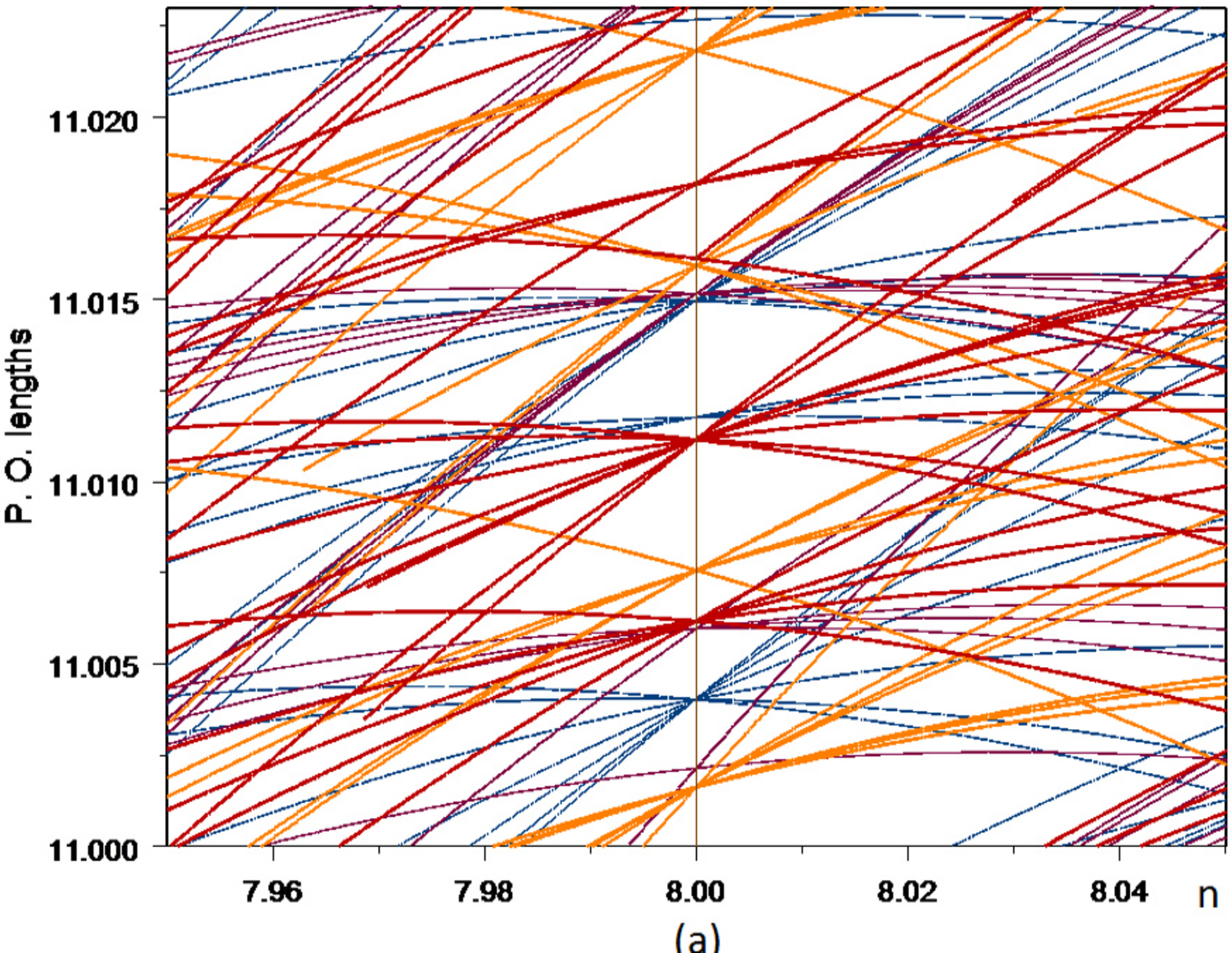} \hfill
\includegraphics[width=0.5\textwidth]{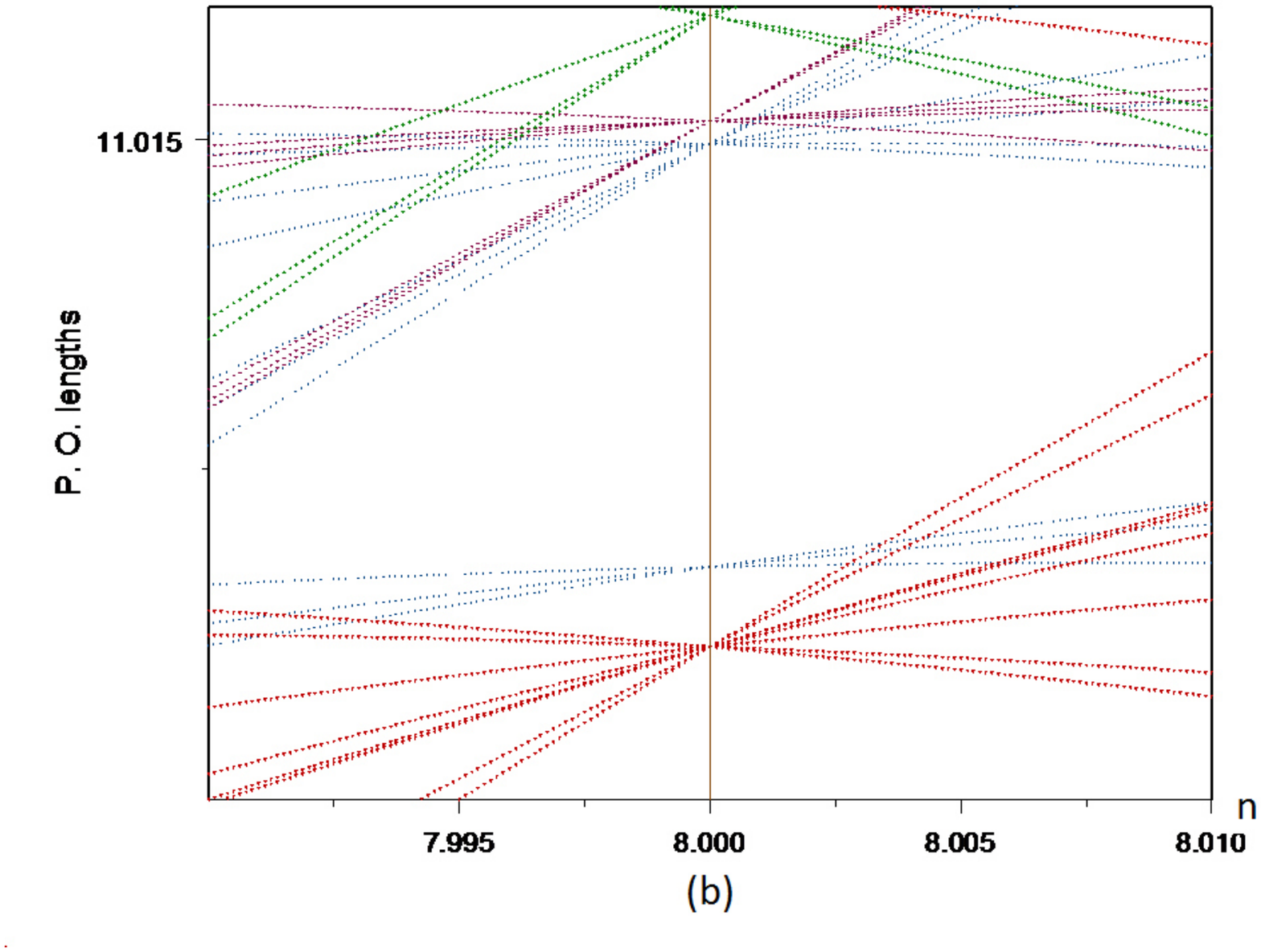} \newline%
{
\parbox[b]{0.3\textwidth}{
} \hfill
\parbox[b]{0.5\textwidth}{
}
\caption{\label{aroundarithm}a)  Stretch of PO  length spectrum of the triangle $(2,3,n)$ with $n$ close to $8$.  The color (red, blue, yellow  or magenta) indicates combination of parities of the reflection numbers $\nu,\nu_L$. Only lines of the same color cross at the arithmetic point $n=8$. b) Blow-up of the length interval $11.011<l<11.015$  }
}
\end{figure}

As an example, we show in Fig.~\ref{aroundarithm}a the periodic orbit lengths  of the triangle with the angles $\left(\pi/2,\pi/3,\pi/n\right),\quad 7.95<n<8.05$. The colors indicate one of the four possible combinations of parities of the reflection numbers $\nu$ and $\nu_M$. At $n=8$ the triangle becomes arithmetic $(2,3,8)$ which is indicated by simultaneous multiple crossings in the plot. Note ``the color segregation'': only lines of the same color (=same parity of $\nu,\nu_M$) are allowed to cross at $n=8$. Fig.\ref{aroundarithm}b shows the blow-up in the length interval $11.011<l<11.015$.

\subsubsection{Right triangles $\left(  2,2k,4k\right)  ,$ $k=2,3,4,6$}

\label{II_2}

Here $\beta=2\alpha=\pi/2k\,\,$. Again we have $\cos\alpha\notin Q\left(
\cos2\alpha,\cos\beta\right)  =Q\left(  2\alpha\right)  $ such that the orbits
with even and odd number of reflections $\nu_{M}$ from the side opposite to
$\beta$ cannot have equal length; the mirror approach doesn't produce an
alternative division since $\cos\beta=\cos2\alpha\in Q\left(  \cos\alpha
,\cos2\beta\right)  =Q\left(  \cos\alpha\right)  =K_{T}$. The statements of
\ref{II_1} can now be repeated: parities of $\nu_{M}$ and $\nu_{L}+\nu_{N}$,
not $\nu_{L},\nu_{N}$ separately, are fixed within the length multiplets. The
quantum triangle is genuinely arithmetic iff the boundary conditions on the
hypotenuse and the leg adjacent to $\beta=\pi/2k$ are the same; we don't know
whether that result was previously reported.

\subsubsection{ Triangle $\left(  2,4,12\right)  $}

\label{II_3}

That right triangle is singled out because the method used up to now fails to
explain the numerically observed division of traces into two types. Indeed
according to the first approach in the case of even $\nu$ the orbit traces
belong to the field $K=Q\left(  \cos\pi/6,\cos\pi/4\right)  $ or to
$K\,\times\cos\pi/12$ depending on parity of $\nu_{M}$. However $\cos\pi
/12=\frac{\sqrt{2}+\sqrt{6}}{4}\in K$ such that $K$ and $K\,\cos\pi/12$
coincide with each other and with the full field $K_{T}$. It follows that
parity of $\nu_{M}$ in the length multiplets is not arithmetically fixed. The
mirror approach is similarly unproductive because $K^{\prime}=Q\left(  \cos
\pi/12\right)  =K_{T}$; consequently parity of $\nu_{N}$ is also not fixed.

In fact, the roles are changed in that triangle: it is parities of $\nu_{L}$
and $\nu_{M}+\nu_{N}$ which are fixed in the multiplets such that the quantum
triangle is genuinely arithmetic if the legs $\ M$ and $N$ have the same
boundary conditions; the boundary condition on the hypothenuse $L$ is
arbitrary. Details can be found in Appendix~\ref{t2_4_12}. That is unique
among the non-symmetric right triangles and probably connected with the fact
that the lengths of the legs $M,N$ are in $2:1$ relation.

\subsubsection{Symmetric triangles whose two equal angles are even fractions
of $\pi$}

\label{II_4}

These include 5 right triangles $\left(  2,2k,2k\right)  ,k=3-6,9,$ and 16
acute ones: $\left(  3,2k,2k\right)  ,\quad k =2,3,4,6;\quad\left(
4,2k,2k\right)  ,\quad k =3,4,8;\quad\left(  5,4,4\right)  ,\left(
5,10,10\right)  ; \newline\left(  6,4,4)\right)  ,\left(  6,12,12\right)
,\left(  6,24,24\right)  ;\quad\left(  9,4,4\right)  ,\left(  9,18,18\right)
;\quad\left(  \infty,4,4\right)  ,\left(  \infty,6,6\right)  . $

These triangles can be desymmetrized by introducing an artificial wall along
the line of symmetry. Periodic orbits of the full triangle can be folded into
its half (\textquotedblleft the fundamental domain\textquotedblright) with the
mirror reflection from the artificial wall. The result is a right arithmetic
triangle; if it is non-symmetric it belongs to one of the types studied above;
if it is still symmetric, one more step of desymmetrization is necessary. It
is important that not all orbits in the desymmetrized triangle but only those
which have  even number of reflections from the fictitious wall, correspond to
the orbits in the original symmetric triangle.

The detailed description is given in Appendix~\ref{appendixSymm}. Here we
mention that if the desymmetrized triangle belongs to the Group I, only two of
the four possible algebraic structures of the trace are realized; therefore
the length spectrum is halved compared with the desymmetrized triangle. On the
other hand, if the desymmetrization result belongs to the Group II the length
spectrum of the symmetric and the desymmetrized triangle coincide; what
differs is the multiplicities in the spectra. In both cases the periodic orbits within a length
multiplet of the original symmetric triangle have fixed parity of the total
number of reflections from the equal sides, not from each of them separately.
Hence the reflection phase in the Gutzwiller expansion has the same value,
modulo $2\pi$, for all orbits within the multiplets iff the boundary
conditions on the symmetric walls coincide; the quantum triangle is then
genuinely arithmetic.

\subsubsection{ Non-symmetric acute triangles}

These are $\left(  3,4,6\right)  ,\left(  3,4,12\right)  ,\left(
3,6,18\right)  ,\left(  3,8,24\right)  ,\left(  3,10,30\right)  $. Here we had
to calculate anew the elementary matrices of reflection and rotation;
investigation and detailed results are given in the Appendix~\ref{piov3}. The
conclusion is that the length multiplets have fixed parity of the number of
reflections from the side $L$ opposite to the angle $\pi/3$ whereas parities
of $\nu_{M}$ and $\nu_{N}$ are not fixed. Therefore the triangles are
genuinely arithmetic if the boundary conditions at the sides adjacent to
$\pi/3$ are the same; that coincides with the group theoretical prediction
mentioned in the Introduction.

\subsection{Group III. Single algebraic type [30 systems]}

The group includes,

\begin{itemize}
\item 9 equilateral triangles $\left(  k,k,k\right)  $, $k=4-9,12,15,\infty;$

\item 21 non-equilateral triangles $\left(  l,m,n\right)  $ in which two or
more of the numbers $l,m,n$ are odd or infinite. These are the right triangles
$\left(  2,3,7\right)  ,$ $\left(  2,3,9\right)  ,$ $\left(  2,3,11\right)
,\left(  2,3,\infty\right)  ,$ $\left(  2,5,5\right)  ,$ $\left(
2,7,7\right)  ,\left(  2,\infty,\infty\right)  $ and the acute ones $\left(
3,3,k\right)  $, $k=4-9,12,15,\infty;$ $\left(  3,5,5\right)  ,\left(
3,\infty,\infty\right)  ,$ $\left(  4,5,5\right)  ,$ $\left(  5,5,10\right)
,$ $\left(  5,5,15\right)  $
\end{itemize}

In these triangles parity of neither $\nu_{L}$ nor $\nu_{M}$ nor $\nu_{N}$ is
fixed within the length multiplet, only that of their sum $\nu$. The quantum
triangle is genuinely arithmetic only if the boundary conditions on all sides
are the same. That conclusion can be confirmed as earlier; e. g., the
equilateral triangles can be desymmetrized resulting in the triangle $\left(
2,k,2k\right)  $ with $\alpha=\pi/2k,\quad\beta=\pi/k$. The wall $M$ adjacent
to $\alpha$ would be a fictitious one such that the number of reflections
$\nu_{M}$ of the orbit folded into the fundamental domain must be even. It was
shown above that the orbit length spectrum of the triangles $\left(
2,k,2k\right)  $ is divided into subspectra with respect to parity of $\nu
_{M}$. However, since odd $\nu_{M}$ are not allowed now, only the division of
the orbit multiplets with respect to parity of $\nu$ remains.

The triangles containing two angles which are odd fractions of $\pi$ are
treated by desymmetrization, if needed, and the usage of (\ref{cospim}).

\section{Conclusion}

Depending on boundary conditions, triangular billiards on the pseudosphere
with one of the 85 \textquotedblleft arithmetic\textquotedblright\ sets of
angles, can have either Poissonian statistics of its energy levels, in spite
of its completely chaotic classical dynamics, or conform to GOE. From the
semiclassical point of view, the peculiar properties of arithmetic systems
result from constructive interference of contributions of an abnormally large
number of periodic orbits with exactly the same length and action. In fact,
not only the length but also the Maslov phase of the orbits needs to be equal;
for a billiard on the pseudosphere that means that orbits within every length
multiplet must have the same total phase gained in reflections from the sides
with Dirichlet boundary condition. Coincidence of the reflection phases can
occur only because of special properties of the periodic orbits of arithmetic
billiards. These properties were the topic of this paper. One such property is
that only orbits with the same parity of the total number of reflections can
have the same length. That rule holds, with statistically insignificant
exceptions, for all arithmetical triangles and guarantees that billiards with
all-Dirichlet sides are arithmetic, never pseudo-arithmetic. Other similar
properties of periodic orbits depend on the triangle in question and concern
the parity of the individual number of reflections from the billiard sides.
These properties have been investigated for all arithmetic trangles and
boundary conditions have been established under which the billiards are
arithmetic and pseudo-arithmetic.

The arithmetic/pseudo-arithmetic division of triangles based on the periodic
orbit analysis must of course give the same results as group-theoretical
considerations. Indeed, the well-known rule of the latter approach that the
sides of a genuinely arithmetic triangle adjacent to the angle $\pi/k$ with
$k$ odd must have the same boundary conditions \cite{Auric95,Ninne95}, 
is  confirmed by the results
of the present paper. The equivalence of the two approaches is almost
self-evident in the case of symmetric triangles. A group-theoretical
derivation of the boundary conditions of arithmeticity for the few remaining
triangles can probably be obtained, too.

Our conclusions on the spectral statistics were based on the diagonal
approximation. However, the distinction between the arithmetic and
pseudo-arithmetic cases must survive even if higher-order terms are taken into
account. The standard off-diagonal contributions of the GOE class stem from
pairs of orbits consisting of the same pieces, some of them time-reversed,
connected in different order and therefore having the same reflection phase 
\cite{Muell05,Muell09},
and only these contributions survive in the pseudo-arithmetic case. In the
arithmetic case the set of contributing pairs must be much more diverse
including pairs of orbits with close action but otherwise unrelated; the
problem has not been investigated so far. Specific off-diagonal effects are
expected in systems like Artin's billiard where exact equality is allowed of
the matrix traces associated with the orbits with even and odd number of
reflections leading to the doublet structure of the length spectrum; the form
factor experiences then a crossover at the Ehrenfest time.. 

\section{Acknowledgement}

The author acknowledges support of the Sonderforschungsbereich SFBTR12
\textquotedblright Symmetries and universality in mesoscopic
systems\textquotedblright\ of the Deutsche Forschungsgemeinschaft. He is
indebted to Fritz Haake for continuous help and useful discussions.

\bibliographystyle{unsrt}
\bibliography{pbraun8}

\appendix

\section{ Triangle $\left(  2,4,12\right)  $}

\label{t2_4_12}

We begin with the even total number of reflections $\nu$ when the orbit matrix
can be represented as a product of the elementary rotation matrices. The orbit
matrices turn out to be of two types,%
\[
R^{I}=%
\begin{pmatrix}
u_{1}+u_{2}3^{1/4}\sqrt{6} & u_{3}+u_{4}\,\,3^{1/4}\sqrt{6}\\
-u_{3}+u_{4}\,\,3^{1/4}\sqrt{6} & u_{1}-u_{2}3^{1/4}\sqrt{6}%
\end{pmatrix}
\]
and%
\[
R^{II}=%
\begin{pmatrix}
u_{1}\sqrt{2}+u_{2}3^{1/4} & u_{3}\sqrt{2}+u_{4}\,\,3^{1/4}\\
-u_{3}\sqrt{2}+u_{4}3^{1/4} & u_{1}\sqrt{2}-u_{2}3^{1/4}%
\end{pmatrix}
\]
where $u_{1/2/3/4}$ denote algebraic numbers belonging to the field $Q\left(
\sqrt{3}\right)  $, i.e., $u=q_{0}+q_{1}\sqrt{3}$. The multiplication rules
for the product of two matrices can be symbolically written,%
\[
R^{I}R^{I}=R^{I},\quad R^{I}R^{II}=R^{II},\quad R^{II}R^{I}=R^{II},\quad
R^{II}R^{II}=R^{I}%
\]

It can be directly checked using (\ref{Elemrot}) with $\alpha=\pi/12,\beta
=\pi/4$ that the rotation matrix $R_{Q}=\Sigma_{M}\Sigma_{N}$ belongs to the
type $R^{I}$ whereas $R_{P}=\Sigma_{N}\Sigma_{L}$ and $R_{O}=\Sigma_{L}%
\Sigma_{M}$ belongs to $R^{II}$. It follows that the type of the orbit matrix
is defined by the total number of elementary rotations $R_{P}$ and $R_{O}$
which is equal to the number of reflections $\nu_{L}$ from the hypotenuse $L$.
Consequently orbits in the length multiplets have definite parity of $\nu_{L}$
and $\nu_{M}+\nu_{N}$; the algebraic structure of the orbit traces is
$q_{0}+q_{1}\sqrt{3}$ or $q_{0}\sqrt{2}+q_{1}\sqrt{6}$ when $\nu_{L}$ is even
or odd, respectively. The quantum triangle will be genuinely arithmetic if the
boundary conditions on the legs $M,N$ coincide.

In the case of odd $\nu$ we represent the orbit code as $\Sigma_{M}\mathbb{R}$
and obtain that $\operatorname*{Tr}\Sigma_{M}R=R_{11}-R_{22}
$ has the algebraic structure $\left(  q_{0}\sqrt{2}+q_{1}\sqrt{6}\right)
3^{1/4}$ if $\nu_{N}$ is even and $\left(  q_{0}+q_{1}\sqrt{3}\right)  3^{1/4}
$ if $\nu_{L}$ is odd.

\section{ Triangles with two equal angles which are even fractions of $\pi$}

\label{appendixSymm}

In Fig. 3 we show a symmetric triangle whose symmetry axis is directed along the
imaginary axis in the Poincar\'{e} half-plane,

\bigskip

\begin{figure}[ptb]
\begin{center}
\includegraphics[
scale=0.3
]{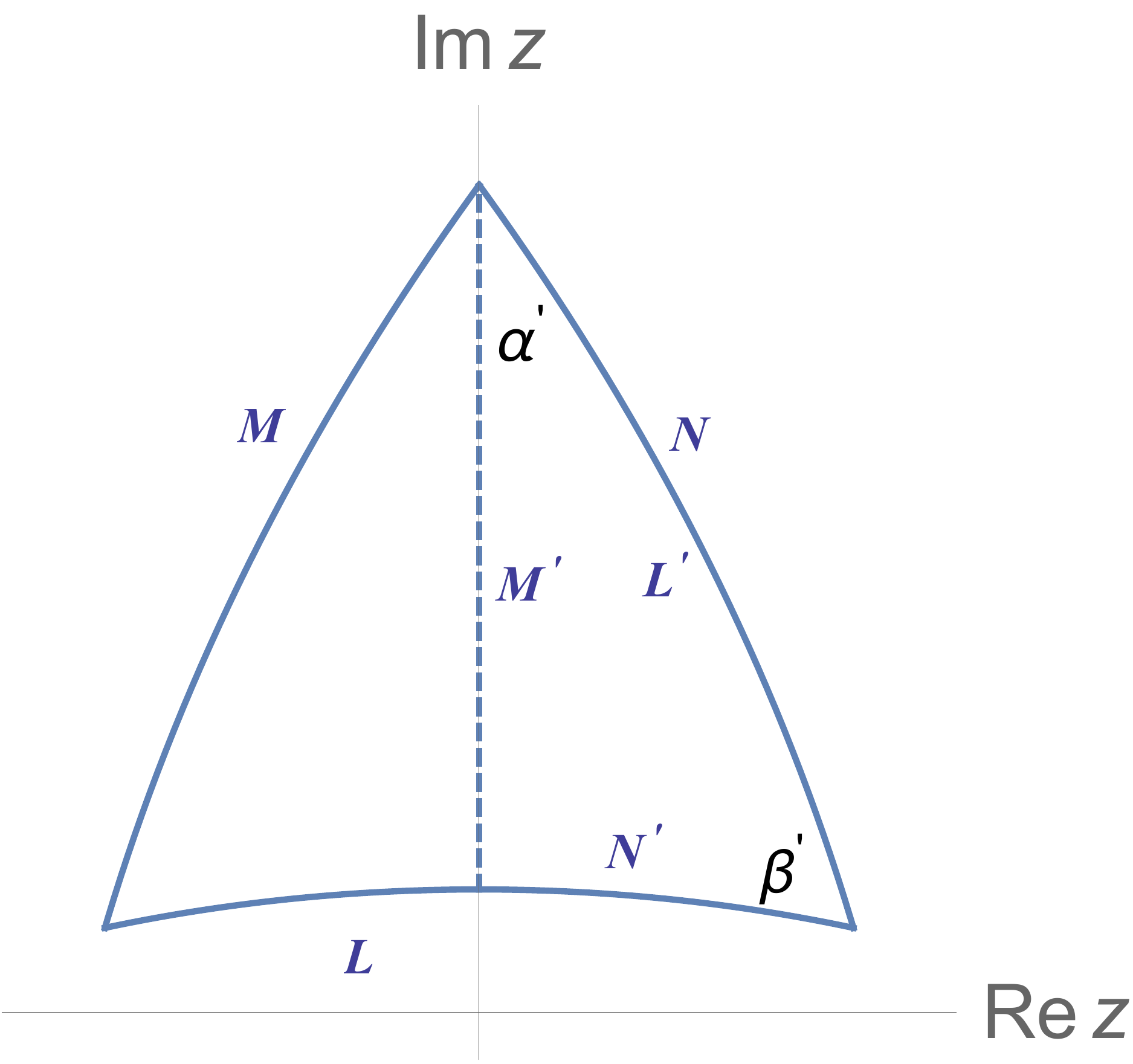}
\end{center}
\caption{Desymmetrization of a triangle with equal sides}%
\label{Fig3}%
\end{figure}Desymmetrizing the problem we draw a fictitious wall denoted
$M^{\prime}$ along the line of symmetry; using notations of our first approach
we denote the side $N$ in the full triangle by $L^{\prime}$ in the
desymmetrized one, and half of the side $L$ by $N^{\prime}$. Then,

\begin{itemize}
\item the number $\nu_{N}^{^{\prime}}$ of reflections from $N^{\prime}$ in the
folded orbit coincides with $\nu_{L}$ in the original one;

\item the side $L^{\prime}$ in the desymmetrized triangle collects reflections
from the sides $M$ and $N$ of the full triangle such that $\nu_{L}^{^{\prime}%
}$ $=$ $\nu_{M}+\nu_{N}$;

\item the number of reflections $\nu_{M}^{^{\prime}}$ is even.
\end{itemize}

Desymmetrizing $\left(  l;2k,2k\right)  $ we obtain the right triangle $\left(
2,m^{\prime},n^{\prime}\right)  $ with even $m^{\prime}=2l$ and $n^\prime=2k$.
Unless $n=k$ it will be non-symmetric and belong to one of the considered
types. The algebraic limitations on parity of the number of reflection
concern $\nu_{L}^{^{\prime}}$ $=$ $\nu_{M}+\nu_{N}$, not $\nu_{M},\nu_{N}$
separately, such that the symmetric sides $M$ and $N$ must have the same
boundary conditions for the quantum triangle to be genuinely arithmetic; this
wass of course to be expected.

Now we give a survey of the results.

A) In 8 triangles a single step of desymmetrization produces a triangle of the
Group I. The matrix traces in that group can be of four types, however only
two of them survive since the number of strikes against the fictitious wall
must be even.%

\begin{tabular}
[c]{ll}%
Original $\left(  l,2k,2k\right)  $ & Desymmetrized $\left(  2,2l,2k\right)
$\\
$\left(  2,6,6\right)  $ & $\left(  2,4,6\right)  $\\
$\left(  2,10,10\right)  $ & $\left(  2,4,10\right)  $\\
$\left(  2,18,18\right)  $ & $\left(  2,4,18\right)  $\\
$\left(  3,4,4\right)  $ & $\left(  2,6,4\right)  $\\
$\left(  3,8,8\right)  $ & $\left(  2,6,8\right)  $\\
$\left(  4,6,6\right)  $ & $\left(  2,8,6\right)  $\\
$\left(  5,4,4\right)  $ & $\left(  2,10,4\right)  $\\
$\left(  9,4,4\right)  $ & $\left(  2,18,4\right)  $%
\end{tabular}

Using our results on Group I it is easy to show that for orbits with even
$\nu=\nu_{N}+\nu_{L}+\nu_{M}$ the traces belong to the field $K=Q\left(
\cos\pi/l,\cos\pi/k\right)  $ if $\nu_{L}$ is even, and to $K\cos\pi/k$ if
$\nu_{L}$ is odd. With odd $\nu$ the traces belong to $\rho\left(  \pi
/2n,\pi/2k\right)  K\cos\pi/k$ and $\rho\left(  \pi/2n,\pi/2k\right)  K$ if
$\nu_{L}$ is even and odd, respectively.

B) In 6 cases the desymmetrized triangle belongs to the Group II:%

\begin{tabular}
[c]{ll}%
Original $\left(  l,2k,2k\right)  $ & Desymmetrized $\left(  2,2l,2k\right)
$\\
$\left(  2,8,8\right)  $ & $\left(  2,4,8\right)  $\\
$\left(  2,12,12\right)  $ & $\left(  2,4,12\right)  $\\
$\left(  3,12,12\right)  $ & $\left(  2,6,12\right)  $\\
$\left(  4,8,8\right)  $ & $\left(  2,8,16\right)  $\\
$\left(  6,4,4\right)  $ & $\left(  2,4,12\right)  $\\
$\left(  6,24,24\right)  $ & $\left(  2,12,24\right)  $\\
$\left(  \infty,4,4\right)  $ & $\left(  2,\infty,4\right)  $\\
$\left(  \infty,6,6\right)  $ & $\left(  2,\infty,6\right)  $%
\end{tabular}

Connection of parity of $\nu_{L}$ and the algebraic properties of the traces
is the same as in A). Note the difference: half of lengths in the orbit length
spectrum of the symmetric triangle of the case A) disappears compared with the
corresponding desymmetrized triangle since two of the four trace algebraic
types are not allowed for the orbits obtained by folding. On the other hand,
in the case B) the length spectra of the symmetrized and desymmetrized systems
coincide, only the multiplicities in the spectra differ.

D) The 5 remaining symmetric triangles need one more stage of desymmetrization%

\begin{tabular}
[c]{lll}%
Original $\left(  l;2k,2k\right)  $ & Once desymmetrized & Second stage,
$\left(  2,4,2k\right)  $\\
$\left(  3,6,6\right)  $ & $\left(  2,6,6\right)  $ & $\left(  2,4,6\right)
$\\
$\left(  4,8,8\right)  $ & $\left(  2,8,8\right)  $ & $\left(  2,4,8\right)
$\\
$\left(  5,10,10\right)  $ & $\left(  2,10,10\right)  $ & $\left(
2,4,10\right)  $\\
$\left(  6,12,12\right)  $ & $\left(  2,12,12\right)  $ & $\left(
2,4,12\right)  $\\
$\left(  9,18,18\right)  $ & $\left(  2,18,18\right)  $ & $\left(
2,4,18\right)  $%
\end{tabular}

\bigskip Denoting the sides of the fundamental domain $\left(  2,4,2k\right)
$ as $L^{\prime\prime},M^{\prime\prime},N^{\prime\prime}$ where the new
fictitious wall $N^{\prime\prime}$ is the bisector of the right angle of the
intermediate triangle $L^{\prime}M^{\prime}N^{\prime}$ and the side
$L^{\prime\prime}$ coincides with $N^{\prime}$, we shall have $\alpha
^{\prime\prime}=\angle\,L^{\prime\prime}M^{\prime\prime}=\pi/2k,$
$\beta^{\prime\prime}=\angle\,N^{\prime\prime}L^{\prime\prime}=\pi/4$.The
number of bounces $\nu_{L^{\prime\prime}}$ of the double-folded orbit against
$L^{\prime\prime}$ is the sum $\nu_{M^{\prime}}+\nu_{N^{\prime}}=$
$\nu_{M^{\prime}}+\nu_{L};$ since $\nu_{M^{\prime}}$ must be even, parity of
$\nu_{L^{\prime\prime}}$ and $\nu_{L}$ is the same. The number of bounces
$\nu_{M^{\prime\prime}}$ from $L^{\prime\prime}$ coincides with $\nu
_{N}^{\prime}=\nu_{N}+\nu_{M},$ i.e. has the same parity as $\nu_{L}.$ The
number of bounces $\nu_{N^{\prime\prime}}$ from the second fictitious wall
$N^{\prime\prime}$ is of course even. We can conclude that for orbits with
even $\nu$ the orbits traces belongs to $K=Q\left(  \cos\pi/k\right)  $ if
$\nu_{L}$ is even, and to $\cos\pi/2k\,\,Q\left(  \cos\pi/k\right)  $ if
$\nu_{L}$ is odd. If $\nu$ is odd the factor $\rho=\sqrt{\cos^{2}\frac{\pi
}{2k}-\frac{1}{2}}$ is to be added and parities of $\nu_{L}$ interchanged.

\section{ Non-symmetric acute triangles}

\label{piov3}

We place the triangle in the Poincar\'e plane such that the side $M=OQ$
is directed along the imaginary axis, the vertex $Q$ at the angle $\pi/3$ has the
coordinate $y_{Q}=1$. The vertex $O$ at the angle $\alpha=\pi/n$ has then the
coordinate
\begin{align*}
y_{O}  &  =\frac{1}{\sqrt{3}\sin\alpha}\left(  \cos\alpha+2\cos\beta
+\rho\right)  ,\\
\rho\left(  \alpha,\beta\right)   &  \equiv\sqrt{1+2\cos2\alpha+2\cos
2\beta+4\cos\alpha\cos\beta};
\end{align*}
the vertex $P$ at the angle $\beta=\pi/m$ is chosen to lie in the first
quadrant of the Poincar\'{e} plane. We denote by $\Sigma_{L/M/N}$ the matrices
of reflections, in particular, as previously $\Sigma_{M}=%
\begin{pmatrix}
1 & 0\\
0 & -1
\end{pmatrix}
$.The matrices of rotation about the vertices read,%
\begin{align*}
R_{O}  &  =\Sigma_{M}\Sigma_{L}=%
\begin{pmatrix}
\cos\alpha & -\frac{1}{\sqrt{2}}\left(  \cos\alpha+2\cos\beta+\rho\right) \\
\frac{1}{\sqrt{2}}\left(  \cos\alpha+2\cos\beta-\rho\right)  & \cos\alpha
\end{pmatrix}
,\\
R_{Q}  &  =\Sigma_{N}\Sigma_{M}=%
\begin{pmatrix}
-\frac{1}{2} & \frac{\sqrt{3}}{2}\\
-\frac{\sqrt{3}}{2} & -\frac{1}{2}%
\end{pmatrix}
;
\end{align*}
the matrix of rotation $R_{P}=\Sigma_{N}\Sigma_{L}$ can be replaced by
$R_{Q}R_{O}$. Any code with even number of reflections $\nu$, can be
written as a sequence of $R_{O}$ and $R_{Q}.$

It will be convenient to denote the triangles here as

$\left(  l,m,n\right)  =\left(  3,6,4\right)  ,\left(  3,12,4\right)  ,\left(
3,18,4\right)  ,\left(  3,24,8\right)  ,\left(  3,30,10\right)  $. The first
triangle stands apart and will be treated separately.

\subsection{Triangle $\left(  3,6,4\right)  $}

Here the logic follows the case of $\left(  2,3,8\right)  $. Substituting
$\alpha=\pi/4,\beta=\pi/6$ we obtain,%
\begin{align*}
R_{O}  &  =%
\begin{pmatrix}
\frac{\sqrt{2}}{2} & -\frac{\left(  6+\sqrt{6}+2r\sqrt{3}\right)  }{6}\\
\frac{6+\sqrt{6}-2r\sqrt{3}}{6} & \frac{\sqrt{2}}{2}%
\end{pmatrix}
,\\
r  &  =\rho\left(  \pi/4,\pi/6\right)  =\sqrt{2+\sqrt{6}}.
\end{align*}
The orbit matrices $R$ are divided into two types which can be written in
terms of the algebraic numbers%
\begin{align*}
u  &  =q_{0}+q_{1}\sqrt{6}\in Q\left(  \sqrt{6}\right)  ,\quad\\
v  &  =q_{0}\sqrt{2}+q_{1}\sqrt{3}\in\sqrt{2}Q\left(  \sqrt{6}\right)
\end{align*}
where $q_{0/1}$ are rationals, as%
\[
R^{I}=%
\begin{pmatrix}
u_{1}+v_{1}r & v_{2}+u_{2}r\\
-v_{2}+u_{2}\,\,r & u_{1}-v_{1}r
\end{pmatrix}
\]
and%
\[
R^{II}=%
\begin{pmatrix}
v_{1}+u_{1}r & u_{2}+v_{2}r\\
-u_{2}+v_{2}\,\,r & v_{1}-u_{1}r
\end{pmatrix}
.
\]

It is easily checked that multiplication by a $R^{II}$-matrix changes the type
of the matrix whereas that by $R^{I}$ doesn't; symbolically, $R^{I}%
R^{I}=R^{I},\quad R^{I}R^{II}=R^{II},\quad R^{II}R^{II}=R^{I}$. The elementary
rotation $R_{Q}$ belongs to the type $I$ with $u_{1}=-1/2,\quad v_{2}=\sqrt
{3}/2,\quad v_{1}=u_{2}=0$; the rotation $R_{O}$ belongs to the type $II$ with
$v_{1}=\sqrt{2}/2,\quad u_{1}=0,\quad u_{2}=-1-\sqrt{6}/6,\quad v_{2}%
=-\sqrt{3}/3.$ It follows that the orbit matrix belongs to the type $I$ (resp.
$II$) if the orbit code contains an even (resp. odd) number of elementary
rotations $R_{O}$; the number of rotations $R_{Q}$ is irrelevant.

 Turning to
the matrix traces and reformulating the result in terms of the number of
reflections $\nu_{L/M/N}\ $ we get that the trace is given by $u\in
Q\left(  \sqrt{6}\right)  $ (resp. $v\in\sqrt{2}Q\left(  \sqrt{6}\right)  $)
for $\nu_{L}$ even (resp. odd). We got thus a refinement of the general
formula (\ref{FieldKt}) which would give $\operatorname*{Tr}R\in Q\left(
\sqrt{2,}\sqrt{3}\right)  $. Since non-zero $u$ and $v$ cannot be equal, the
orbit length spectrum falls into two subspectra; the orbits within each
degenerate multiplet have the same parity of $\nu_{L}$ and $\nu_{M}+\nu_{N}$
but not of $\nu_{M},\nu_{N}$ individually. Consequently, in the quantum
problem the statistics will be Poissonian and the triangle genuinely
arithmetic iff the boundary conditions at the sides $M,N$ are the same (both
Neumann or both Dirichlet).

In the case of odd $\nu$ , i.e, the inverse hyperbolic orbits, we represent
the orbit code as $\Sigma_{M}\mathbb{R}$ and obtain the trace of the orbit
matrix as the difference of the diagonal elements of the rotational body. It
follows that the trace belongs to $rQ\left(  \sqrt{6}\right)  $ if $\nu_{L}$
is even, otherwise it is $r\,\,\sqrt{2}Q\left(  \sqrt{6}\right)  $ These two
sets don't have common non-zero elements such that $\nu_{L},\nu_{M}+\nu_{N}$
have fixed parity within the length multiplets; all conclusions for the even
$\nu$ remain thus in force. Note that $r=\sqrt{2+\sqrt{6}}$ does not belong to
$Q\left(  \sqrt{2},\sqrt{3}\right)  $, i.e., traces in the odd case belong to
an extension of the field $K_{T}$.

\subsection{Triangles $\left(  3,12,4\right)  ,\left(  3,18,6\right)  ,\left(
3,24,8\right)  ,\left(  3,30,10\right)  $}

In the remaining non-right triangles we have $m=3n$ such that $\alpha=3\beta$.
We start with the case of even $\nu$ when the code is a product of elementary
rotations. The matrices of the orbit code fall into two types; introducing
$y=2\cos\beta$ we can write them as%
\begin{equation}
R^{I}=%
\begin{pmatrix}
\frac{P_{1}^{I}\left(  y^{2}\right)  +\sqrt{y^{2}-3}Q_{1}^{I}\left(
y^{2}\right)  }{2} & \frac{P_{2}^{I}\left(  y^{2}\right)  +y\sqrt{y^{2}%
-3}Q_{2}^{I}\left(  y^{2}\right)  }{2\sqrt{3}}\\
\frac{-P_{2}^{I}\left(  y^{2}\right)  +y\sqrt{y^{2}-3}Q_{2}^{I}\left(
y^{2}\right)  }{2\sqrt{3}} & \frac{P_{1}^{I}\left(  y^{2}\right)  -\sqrt
{y^{2}-3}Q_{1}^{I}\left(  y^{2}\right)  }{2}%
\end{pmatrix}
\label{RI}%
\end{equation}
and%
\begin{equation}
R^{II}=%
\begin{pmatrix}
\frac{yP_{1}^{II}\left(  y^{2}\right)  +\sqrt{y^{2}-3}Q_{1}^{II}\left(
y^{2}\right)  }{2} & \frac{yP_{2}^{II}\left(  y^{2}\right)  +\sqrt{y^{2}%
-3}Q_{2}^{II}\left(  y^{2}\right)  }{2\sqrt{3}}\\
\frac{-yP_{2}^{II}\left(  y^{2}\right)  +\sqrt{y^{2}-3}Q_{2}^{II}\left(
y^{2}\right)  }{2\sqrt{3}} & \frac{yP_{1}^{II}\left(  y^{2}\right)
-\sqrt{y^{2}-3}Q_{1}^{II}\left(  y^{2}\right)  }{2}%
\end{pmatrix}
. \label{RII}%
\end{equation}
Here $P_{1,2}^{I,II}\left(  y^{2}\right)  ,Q_{1,2}^{I,II}\left(  y^{2}\right)
$ are polynomials of $y^{2}$ with rational coefficients. Multiplication rules
for matrices belonging to the two types are the same as in the preceding
section, $R^{I}R^{I}=R^{I},\quad R^{I}R^{II}=R^{II},\quad R^{II}R^{II}=R^{I}$.
The elementary rotation matrix $R_{Q}$ belongs to the type $I$ with $P_{1}%
^{I}=-1,\quad P_{2}^{I}=3,\quad Q_{I}^{1}=Q_{2}^{I}=0;$ the matrix $R_{O}$ is
of the type $II$ with $P_{1}^{II}=y^{2}-3,\quad Q_{1}^{II}=0,\quad P_{2}%
^{II}=1-y^{2},\quad Q_{2}^{II}=2\left(  1-y^{2}\right)  $.

Using the multiplication rules we can prove (\ref{RI}),(\ref{RII}) and obtain
that the code matrix belongs to the type I (resp. II) if the code contains an
even (resp. odd) number of rotations $R_{O}$ equal to the number of
reflections $\nu_{L}$ . The matrix traces are given by $P_{1}^{I}\left(
y^{2}\right)  \in Q\left(  \cos2\beta\right)  $ or $yP_{1}^{II}\left(
y^{2}\right)  \in\cos\beta\,Q\left(  \cos2\beta\right)  $, respectively; since
$\beta=\pi/m$ with $m$ even, these two sets do not have non-zero common
elements, i. e., the orbits with $\nu_{L}$ of different parity cannot have
equal lengths. Consequently, parity of $\nu_{L}$, same as that of $\nu_{M}%
+\nu_{N}$, is fixed within each length multiplet whereas individual parities
of $\nu_{M},\nu_{N}$ are not. Therefore the quantum triangle is genuinely
arithmetic iff the boundary conditions on $M$ and $N$ are both Dirichlet or
both Neumann.

The case of odd $\nu$ is treated similar to $\left(  3,6,4\right)  $; the
traces belong to $\rho Q\left(  \cos2\beta\right)  $ or to $\rho Q\left(
\cos2\beta\right)  \,\,\cos\beta\,\,$ with $\rho=\sqrt{4\cos^{2}\beta-3}$if
$\nu_{L}$ is even or odd, respectively.

\end{document}